\documentclass[aps,prl,twocolumn,showpacs,superscriptaddress]{revtex4}

\usepackage{graphicx}

\begin{document}

\title{Resonant Photovoltaic Effect in Surface State Electrons on Liquid Helium}

\author{Denis Konstantinov}
\affiliation{Okinawa Institute of Science and Technology, Tancha 1919-1, Okinawa 904-0412, Japan}
\affiliation{Low Temperature Physics Laboratory, RIKEN, Hirosawa 2-1, Wako 351-0198, Japan}
\author{A.D. Chepelianskii}
\affiliation{Low Temperature Physics Laboratory, RIKEN, Hirosawa 2-1, Wako 351-0198, Japan}
\affiliation{LPS, Univ. Paris-Sud, CNRS, UMR 8502, F-91405 Orsay Cedex, France}
\author{Kimitoshi Kono}
\affiliation{Low Temperature Physics Laboratory, RIKEN, Hirosawa 2-1, Wako 351-0198, Japan}

\begin{abstract}
We observed an ultra-strong photovoltaic effect induced by resonant intersubband absorption of microwaves in a two-dimensional electrons system on the surface of liquid helium. The effect emerges in the regime of microwave-induced vanishing of dissipative conductance, $\sigma_{xx}\rightarrow 0$, reported previously [D. Konstantinov and K. Kono: Phys. Rev. Lett. \textbf{105} (2010) 226801)] and is characterized by a nonequilibrium spatial distribution of electrons in the confining electrostatic potential. The electrostatic energy acquired by an electron exceeds other relevant energies by several orders of magnitude.
\end{abstract}

\pacs{73.20.At, 73.21.-b, 72.20.My, 78.70.Gq}

\maketitle

\begin{figure}[t]
\centering
\includegraphics[width=8.5cm]{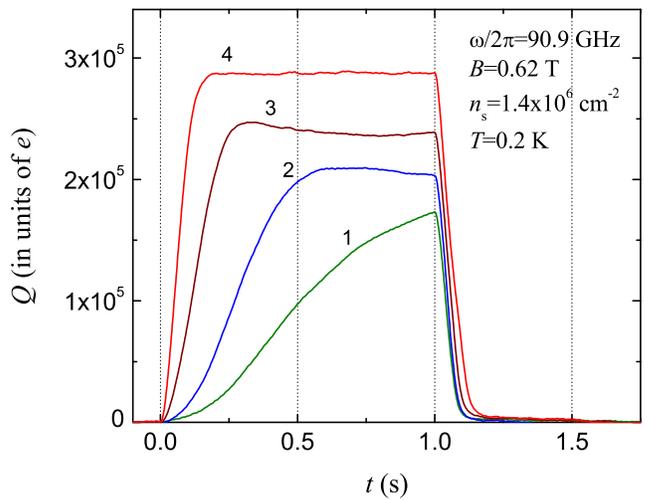}
\caption{\label{fig:5}(color online) Cumulative charge in electrode C$_1$ at $B=0.62$~T for four power levels -24 (curve 1), -20 (curve 2), -10 dBm (curve 3), and -3 dBm (curve 4) of the input microwave power. The power is switched on (off) at $t$=0 (1.0 s).} 
\end{figure}

A nondegenerate two-dimensional electron system can be formed on the surface of liquid helium~\cite{Andrei, Monarkha_book}. Here, surface state subbands with energies $\epsilon_n$ $(n=1,2,..)$ appear owing to the attractive image force, the repulsive surface barrier, and an electric field $E_{\perp}$ applied perpendicular to the surface. Below 1~K, almost all electrons are frozen into the lowest subband forming an equipotential 2D charge layer confined on helium surface. It is a strongly-correlated system of particles interacting via unscreened Coulomb interaction. A recent work reported an effect of vanishing conductance in this system under resonant excitation by microwaves~\cite{Konstantinov2010}. In the experiment, the inter-subband $n=1\rightarrow 2$ transition is excited using radiation with angular frequency $\omega$ such that $\hbar\omega\approx \epsilon_2 -\epsilon_1$. Under such conditions, the longitudinal conductivity $\sigma_{xx}$, measured as {\it a time-averaged response} using a Corbino disc, oscillates upon varying the perpendicular magnetic field $B$, showing a sequence of minima shifted to lower values with respect to $B$ satisfying the relation $\omega/\omega_c=l$, where $\omega_c=eB/m$ is the cyclotron frequency and $l=4,5,..$ (the lowest $l$ was limited by $B<0.85$~T employed in the experiment). At low $l$, the conductivity drops abruptly to zero. In accordance with the standard tensor relation, the vanishing conductivity $\sigma_{xx}\rightarrow 0$ at the minima corresponds to vanishing resistivity $\rho_{xx}$ suggesting connection with radiation-induced zero-resistance states (ZRS) of degenerate 2D electron gas in high mobility GaAs/AlGaAs heterostructures~\cite{Mani2002,Zudov2003}.
\newline
\indent In this work, we study {\it a transient response} of the electron system upon irradiation in the regime of vanishing conductance. We show that the radiation causes strongly nonequilibrium distribution of electrons in the confining electrostatic potential produced by the surrounding metallic electrodes. This corresponds to an increase of the electrostatic potential energy of an electron of the order electron volt, which exceeds any relevant energies, such as for example the average kinetic energy of the electrons, by orders of magnitude. The effect must have strong connection with the zero-resistance states.
\newline
\begin{figure}[b]
\centering
\includegraphics[width=8.5cm]{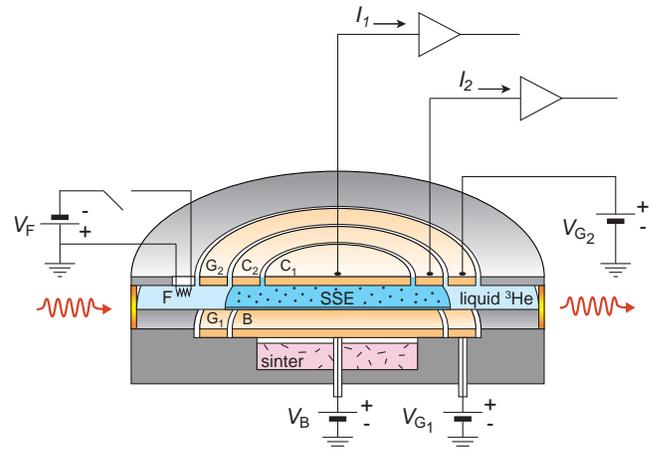}
\caption{\label{fig:1}(color online). Schematic diagram of the experimental method. Detailed description is provided in the text.} 
\end{figure}
\indent Figure~\ref{fig:1} shows a schematic diagram of the experimental apparatus. Liquid $^3$He is condensed into a cell having flat cylindrical shape, which contains a sintered silver heat exchanger to ensure sufficient heat contact between the liquid and the cell body. The liquid level is set midway between two parallel plates, each having a radius of 13~mm, separated by $d=2.6$~mm. The bottom plate is divided into electrodes B and G$_1$ by a gap at a distance of 10~mm from the center of the plate, while the top plate is divided into electrodes C$_1$, C$_2$ and G$_2$ by two gaps at distances of 7 and 10~mm from the center. Each of the three gaps has a width of 0.2~mm. Electrodes C$_1$ and C$_2$ form a Corbino disk (dc-grounded), from which $\sigma_{xx}$ can be determined as described previously~\cite{Konstantinov2010}. Electrons are produced by briefly heating a tungsten filament F, while a positive bias is applied to the electrode B and all other electrodes are grounded. The nearly uniform areal density of electrons, which form a circular pool on the surface beneath the electrode, is determined by the condition of complete screening of electric field above the charged surface of liquid~\cite{Konstantinov2010}. A typical bias of $V_B=0.4$~V used in this experiment corresponds to the areal density of electrons $n_s\approx 1.7\times 10^6$~cm$^{-2}$. However, this number can vary due to accidental discharge of electrons or change of the electrical biases applied to surrounding electrodes. A careful calibration procedure, that allows for the total number of electrons on the surface to be determined, is described further in the text.
\newline
\indent The positively biased electrodes provide a neutralizing background, which counteracts the Coulomb repulsion between electrons and prevents them to spread away. Within a conducting charged layer, the in-plane electric field is almost zero. Below 1 K, the average thermal kinetic energy of an electron does not exceed $10^{-4}$~eV. Therefore, a built-in electric field, which compensates the diffusion current of electrons in a density gradient, can be neglected, and the equilibrium spatial distribution of 2D electrons can be obtained numerically assuming constant electrical potential across the entire charged layer~\cite{Wilen1988}. The equilibrium density profiles for different electrode biases will be discussed later.
\newline
\indent Electrons are tuned for the intersubband $n=1\rightarrow 2$ resonance with the applied microwaves by adjusting the electrode bias $V_{\rm B}$, which changes the electric field $E_{\perp}$ and shifts the subband energies because of the Stark effect. For microwaves of frequency $\omega/2\pi=90.9$~GHz employed in this experiment, the resonance is centered at $V_{\rm B}=6.4$~V. Usually, the guard electrodes G$_1$ and G$_2$ are kept at zero potential. Previously, the photo-induced change in $\sigma_{xx}$ was observed upon cw irradiation and measuring a time-averaged response of the electrons to a driving in-plane electric field~\cite{Konstantinov2010}. The present work aims at detecting the transient photoresponse of surface electrons in the absence of such a field. For this purpose, the incident microwave power is pulse modulated using a low-frequency (0.5-6 Hz) square waveform, and the current signals $I_1$ and $I_2$ induced in electrodes C$_1$ and C$_2$, respectively, are recorded using a digital storage oscilloscope following the current preamplifiers (100~Hz bandwidth). Alternatively, these currents or their difference $\Delta I=I_1-I_2$ are measured by a lock-in amplifier using the microwave modulating signal as a reference.
\newline
\begin{figure}[b]
\centering
\includegraphics[width=8.5cm]{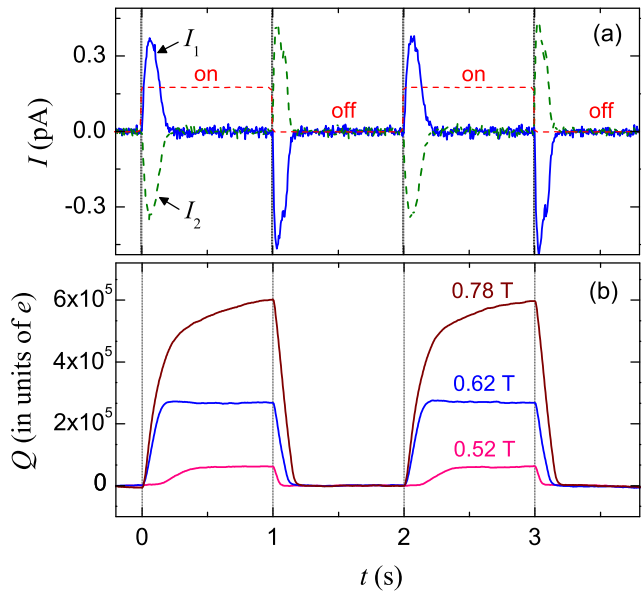}
\caption{\label{fig:2}(color online) (a) Transient signals of photocurrents $I_1$ (solid line, blue) and $I_2$ (dashed line, green) induced in electrodes C$_1$ and C$_2$, respectively, by the flow of the surface charge at $T=0.2$~K and $B=0.62$~T. Short dashed line (red) is a square waveform, which switches the MW source on (off) at a high (low) signal level. (b) Cumulative charge $Q$ obtained by integrating the current $I_1$ at three values of $B$ corresponding to $l=4$ (0.78 T), $l=5$ (0.62 T), and $l=6$ (0.52 T) conductance minima.} 
\end{figure}
\indent Figure~\ref{fig:2}(a) shows an example of the transient current signals $I_1$ and $I_2$ obtained at $T=0.2$~K and $B=0.62$~T. The value of $B$ corresponds to the $l=5$ conductance minimum. In order to improve signal-to-noise ratio, the traces shown in the figure were obtained by averaging over about 500 sweeps. For reference, the modulation of the microwave power by the square waveform is shown in Fig.~\ref{fig:2}(a) by a dashed line. First of all, note that the current signals $I_1$ and $I_2$ are comparable in magnitude and have opposite signs. From this result, we obtain the following qualitative picture. Upon switching the power on, the electrons are pulled towards the edge of the electron pool, causing the depletion (accumulation) of the charge in the central (peripheral) region of the pool. Correspondingly, the positive (negative) current is induced in electrode C$_1$ (C$_2$) by the flow of the image charge. The surface charge flows until a new spatial distribution of 2D electrons in the unchanged confining electrostatic potential is established, after which the currents $I_1$ and $I_2$ becomes zero. Correspondingly, the displacement of electrons with respect to the neutralizing background induces variation in the electrostatic potential energy of an electron, and therefore a non-zero build-in electric field, to develop in the charged layer. Upon switching the power off, the displaced surface charge flows back to restore the equilibrium distribution of electrons. Correspondingly, a negative (positive) current of the image charge is induced in C$_1$ (C$_2$).
\newline
\begin{figure}[b]
\centering
\includegraphics[width=8.5cm]{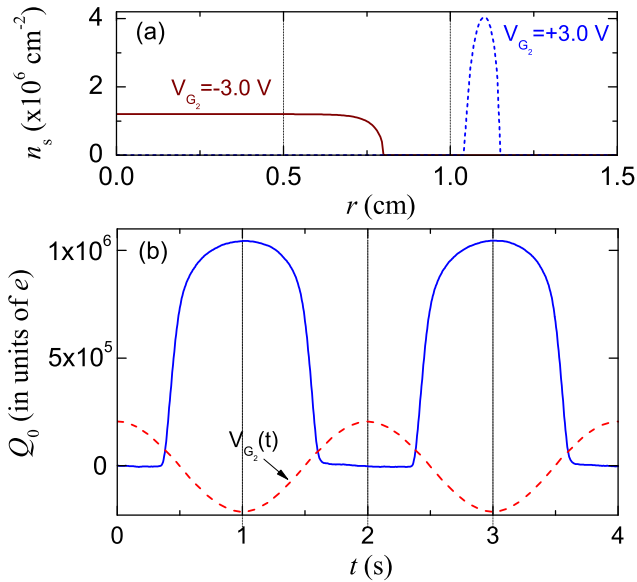}
\caption{\label{fig:3}(color online) (a) Areal density of electrons versus distance from center of the cell obtained numerically for $V_{\rm B}=V_{\rm G_1}=6.4$~V, $V_{\rm G_2}=-3.0$~V (solid line, brown) and $V_{\rm B}=V_{\rm G_1}=6.4$~V, $V_{\rm G_2}=+3.0$~V (dashed line, blue). The total number of electrons is $2\times 10^6$. (b) Positive image charge induced at the electrode ${\rm C_1}$ by the surface electrons in the dark when an ac (0.5~Hz) potential of amplitude 3~V (dashed line) is applied to the electrode ${\rm G_2}$.} 
\end{figure}
\indent Next, we estimate the amount of the charge displaced under irradiation. The cumulative charge $Q$ flowing from, for example, the electrode C$_2$ is obtained by integrating the measured current $I_1$. $Q$ (in units of the elementary charge $e>0$) is shown in Fig.~\ref{fig:2}(b) for three values of $B$ corresponding to the conductance minima $l=4$, 5, and 6. To make comparison of this quantity with the total image charge induced on electrode C$_1$ in the dark, the latter was experimentally determined from the following procedure. At $V_{\rm B}=V_{\rm G_1}=6.4$~V, we apply an ac (0.5~Hz) potential of the amplitude 3.0~V to the electrode G$_2$. When $V_{\rm G_2}=+3.0$~V, the surface charge is pulled out from the center to form a ring beneath the electrode G$_1$, as confirmed numerically by calculating the charge profile as shown in Fig.~\ref{fig:3}(a). As $V_{\rm G_2}$ decreases to -3.0~V, the charge is pushed back into the region beneath the electrode C$_1$ (see Fig.~\ref{fig:3}(a)) inducing the negative current through the electrode C$_1$. This current is measured and integrated to obtain the positive charge $Q_0$ in the electrode C$_1$ induced by the surface electrons. The induced charge $Q_0$ is shown in Fig.~\ref{fig:3}(b). We find that a total image charge of about $1.0\times 10^6 e$ is induced on the electrode C$_1$. The comparison of this value with $Q$ in Fig.~\ref{fig:2}(b) shows that a very large fraction (more than 50$\%$) of the surface charge can be displaced upon irradiation. From this procedure and the numerial calculations we find the total number of electrons $2\times 10^6$ and the areal density of electrons $n_s=1.4\times 10^6$~cm$^{-2}$ in the dark at $V_{\rm B}=6.4$~V and $V_{\rm G_1}=V_{\rm G_2}=0$.
\newline
\begin{figure}[b]
\centering
\includegraphics[width=8.5cm]{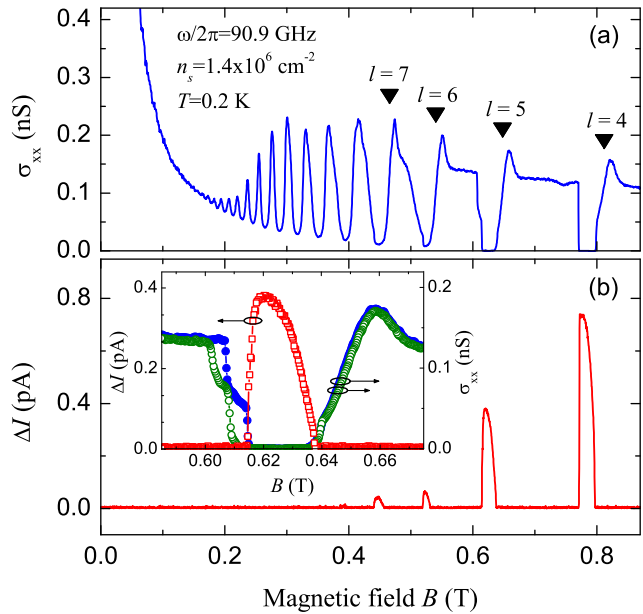}
\caption{\label{fig:4}(color online) (a) Microwave-induced magnetooscillations of electrons obtained for $n_s=1.4\times 10^6$~cm$^{-2}$ at $T=0.2$~K under CW irradiation. (b) Current signal $\Delta I$ obtained under 100$\%$ modulation of the microwaves, and for the same $n_s$ and $T$. Black triangles indicate the values of $B$ such that $\omega/\omega_c=l$. In the inset: $\Delta I$ (left axis) in the range of $B$. For comparison, $\sigma_{xx}$ (right axis) is shown for upward (solid circles) and downward (open circles) sweeps of $B$.} 
\end{figure}
\indent The displacement of negative charged electrons induces a nonzero in-plane electric field to develop across the system. It is straightforward to estimate the radially symmetric built-up difference of electrical potential across the charge layer. For the geometry shown in Fig.~\ref{fig:1}, the electrical potential of the charge layer is given by $V_e=0.5(V_B-e n_sd/(2\varepsilon_0))$, where $d$ is the distance between top and bottom plates and we neglect small deviation of the dielectric constant of liquid helium from unity. Thus, the displacement of 50$\%$ of electrons corresponds to the potential difference between the central and peripheral parts $\Delta V_e=3en_sd/(8\varepsilon_0)$. For the areal density estimated above this gives $\Delta V_e\approx 0.3$~V. This result is confirmed by the numerical calculations and corresponds to the increase in electrical potential energy of a single electron exceeding other relevant energy scales such as, for example, the intersubband energy difference or average thermal kinetic energy of an electron, by several orders of magnitude.
\newline 
\indent Figure~\ref{fig:4} shows a comparison between $\sigma_{xx}$ measured under cw irradiation and $\Delta I$ recorded under 100$\%$ modulation of the incident microwave power. Both sets of data are obtained upon increasing the magnetic field $B$ at $T=0.2$~K and at the same level of microwave power. A nonzero signal $\Delta I$ is observed only in the intervals of $B$ near the conductance minima corresponding to $l=4$, 5, 6, and 7. As described above, this signal is due to the currents induced in electrodes C$_1$ and C$_2$ by the displacement of the surface charge. No signal is observed when electrons are tuned away from the intersubband resonance by changing the electrical bias $V_{\rm B}$. Simultaneously, the detuning results in the complete disappearance of the magnetooscillations and zero-conductance states~\cite{Konstantinov2010}. Importantly, the disappearance of $\Delta I$ in the electron system tuned away from the resonance, as well as when the surface is completely discharged by reversing the sign of $V_{\rm B}$, confirms that the current is not produced in the conducting electrodes by, for example, microwave-induced heating, vibrations, or other effects.
\newline
\indent The inset of Fig.~\ref{fig:4}(b) shows the signal $\Delta I$ in a narrow range of $B$ near the $l=5$ conductance minimum. The signal emerges abruptly upon slowly increasing $B$. This feature is consistent with the abrupt drop of $\sigma_{xx}$ to zero recorded upon the upward sweep of $B$ (solid circles). The abrupt change of $\sigma_{xx}$ and $\Delta I$ is an indication of instability of the electron system in ZRS regime due to, for example, the resonant overheating of electrons by microwave radiation~\cite{Konstantinov2009}. Upon the downward sweep of $B$, $\sigma_{xx}$ (open circles) exhibits hysteresis. Such hysteresis is a feature of a metastable state coexisting with the global stable state of electron system. The coexistence regime can be observed only upon continuous irradiation by microwaves. When the power is repeatedly switched on and off, the system remains in the global stable state corresponding to equilibrium spatial distribution of charge. Accordingly, the variation of $\Delta I$ in Fig.~\ref{fig:4}(b) does not show dependence on the direction of sweeping.
\newline
\indent The kinetics of the radiation-induced displacement of surface charge strongly depends on different parameters such as the microwave power, the density of electrons, and the index $l$. The cumulative charge measured for $n_s=1.4\times 10^6$~cm$^{-2}$, $B=0.62$~T, and at four different power levels is shown in Figure~\ref{fig:5}. For the fully screened electron system in our experiment, the typical time required to recover the equilibrium spatial distribution, the relaxation time $\tau_{\rm d}$, can be estimated as $C/\sigma_{xx}$, where $C$ is the capacitance between the charged layer and the surrounding electrodes. For $C\sim 1$~pF and $\sigma_{xx}\sim 10^{-10}$~S, this gives $\tau_{\rm d}\sim 10^{-2}$~s, which is in agreement with the fall time of $Q$ upon switching the microwaves off, as shown in Fig.~\ref{fig:4}. In contrast, the time required to establish the redistribution of electrons across the layer upon switching the power on, as can be seen from the rising edge of $Q$, can exceed 1~s at sufficiently low powers. Even at high powers, the build-up of charge is slowed by either increasing $n_s$ or choosing the conductance minima of a high index $l$. Remarkably, a time delay of up to 0.1~s between the application of microwaves and the onset of the charge motion can be observed, as shown, for example, by the data labeled as 0.52~T in Fig.~\ref{fig:2}(b). This slow redistribution kinetics is in extreme contras with a much faster rate of microwave-induced inter-subband transitions on the order $10^{6}$-$10^{7}$~s$^{-1}$.  
\newline  
\indent Finally, we discuss the possible explanations for the observed photovoltaic effect. Radiation can exert an optical force on a charged particle, as is used for the cooling and trapping of atoms and ions~\cite{Metcalf_book}. This force can be estimated as the expectation value $-\langle \partial \hat{H}/ \partial x \rangle$, where $\hat{H}$ describes the interaction of the particle with the radiation field and $x$ is the distance along the surface. Our estimations show that the optical force is too weak to directly account for the displacement of charge reported here. The appearance of this effect at the minima of the conductance oscillations indicates the inherent relationship between the redistribution of charge and $\sigma_{xx}$. In addition, an important question is the relation of the effect reported here to photovoltaic effects, which accompany the microwave-induced resistance oscillations in GaAs/AlGaAs heterostructures~\cite{Bykov2008,Willett2004,Dorozhkin2009,Bykov2010}. In heterostructures, the observed oscillations of photocurrent and photovoltage were interpreted in terms of built-in electric fields due to band bending at contacts~\cite{Dorozhkin2009,Dmitriev2009}. Clearly, we do not have such fields in our system where the surface charge layer is maintained at a constant electrical potential at least without irradiation by microwaves. The microwave field gradient in the near-contact region can generate a ponderomotive force acting on 2D electrons in heterostructures, which may be a possible explanation for the ZRS in this system~\cite{Mikhailov2011}. Such an effect is unlikely to occur in our system because the electrons are not in contact with the conducting electrodes. Recently, it was proposed that microwave-induced oscillations of $\sigma_{xx}$ in surface state electrons on liquid helium originate from the intersubband scattering of electrons under the condition of nonequilibrium filling of the excited subband~\cite{Monarkha2011,Monarkha2012}. This mechanism can result in an absolute negative $\sigma_{xx}$, causing current instability and the formation of a steady azimuthal current $j_0$, such that $\sigma_{xx}(j_0)=0$~\cite{Andreev2003}. In a perpendicular magnetic field, the steady configuration would require a radial Hall field, $E_H=j_0/\sigma_H$, where $\sigma_H=n_se/B$. The only way to produce such a field is to displace electrons radially. The measured $E_H$ of the order 1~V/cm would correspond to the current density $j_0\sim$~1nA/cm. This value must be compared with predictions of nonlinear transport theory which is not available at present. Finally, the microwave stabilization of edge trajectories, which was proposed recently to explain ZRS in heterostructures, naturally gives the correct direction of the charge displacement~\cite{Chepelianskii2009}. The creation of ballistic channels at the sample edges due to microwave irradiation combined with a pumping mechanism through the inter-subband excitation might lead to an accumulation of charge at the edges. This possibility, as well as other outlined scenarios, require thorough theoretical investigation.
\newline
\indent In summary, we studied the photoresponse of a circular pool of electrons on the surface of liquid helium subjected to a perpendicular magnetic field. We found out that under the resonant intersubband-absorption of microwaves, a large fraction of electrons is displaced from the center of the pool towards the edge producing a strongly nonequilibrium distribution of charge. The effect emerges at minima of the microwave-induced magneto-oscillations and suggests close connection with the states of vanishing resistance.
\newline
\indent We thank H. Bouchiat and D.L. Shepelyansky for helpful discussions. This work was supported in part by Grant-in-Aids for Scientific Research from MEXT. D.K. acknowledges an Incentive Research Grant from RIKEN and internal grant from the Okinawa Institute of Science and Technology.

\end{document}